\documentclass[letterpaper, 10 pt, conference]{ieeeconf}  % Comment this line out
\IEEEoverridecommandlockouts                              % This command is only
\usepackage{amsfonts}
\usepackage{amsmath}
\usepackage{graphicx}
\usepackage[breaklinks=true]{hyperref}
\usepackage{caption,setspace, subcaption}
\usepackage{etoolbox}
\makeatletter
\patchcmd{\@makecaption}
  {\scshape}
  {}
  {}
  {}
\makeatletter
\patchcmd{\@makecaption}
  {\\}
  {.\ }
  {}
  {}
\makeatother

\graphicspath{ {./images/} }

\usepackage{array}
\newcolumntype{P}[1]{>{\centering\arraybackslash}p{#1}}

\usepackage[utf8]{inputenc}
\usepackage[english]{babel}
\usepackage{csquotes}

\usepackage[backend=biber, style=nature, sorting=ynt]{biblatex}

\title{\LARGE \bf
MammoGANesis: Controlled Generation of High-Resolution Mammograms for Radiology Education
}

\author{Cyril Zakka$^{1}$, Ghida Saheb$^{2}$, Elie Najem$^{2}$, Ghina Berjawi$^{2}$% stops a space
\thanks{$^{1}$\textit{Faculty of Medicine, American University of Beirut, Lebanon}}%
\thanks{$^{2}$\textit{Department of Diagnostic Radiology, American University of Beirut Medical Center, Lebanon}}%
}

\begin{document}

\maketitle
\thispagestyle{empty}
\pagestyle{empty}

%%%%%%%%%%%%%%%%%%%%%%%%%%%%%%%%%%%%%%%%%%%%%%%%%%%%%%%%%%%%%%%%%%%%%%%%%%%%%%%%
\begin{abstract}
During their formative years, radiology trainees are required to interpret hundreds of mammograms per month, with the objective of becoming apt at discerning the subtle patterns differentiating benign from malignant lesions. Unfortunately, medico-legal and technical hurdles make it difficult to access and query medical images for training.

In this paper we train a generative adversarial network (GAN) to synthesize 512 x 512 high-resolution mammograms. The resulting model leads to the unsupervised separation of high-level features (e.g. the standard mammography views and the nature of the breast lesions), with stochastic variation in the generated images (e.g. breast adipose tissue, calcification), enabling user-controlled global and local attribute-editing of the synthesized images.

We demonstrate the model's ability to generate anatomically and medically relevant mammograms by achieving an average AUC of 0.54 in a double-blind study on four expert mammography radiologists to distinguish between generated and real images, ascribing to the high visual quality of the synthesized and edited mammograms, and to their potential use in advancing and facilitating medical education.

\end{abstract}
%%%%%%%%%%%%%%%%%
\section{Introduction}

Over the course of their medical education, radiology trainees are required to interpret hundreds of images per month in order to obtain basic competency in visual diagnosis \cite{Wang2012}. Performance on these interpretations improves with increasing exposure to mammograms, with higher detection rates of cancers and lower unnecessary work-ups noted in radiologists with additional fellowship training and targeted medical education \cite{Miglioretti2009}. 

Nevertheless, working within the context of medical records and images poses unique legal and technical challenges that can prove to be real barriers for medical education and research \cite{pmid29618526}. Clinical data is often heterogeneous and messy \cite{DBLP:journals/corr/abs-1806-00388}, and often unamenable to simple querying. Despite the availability of structured data (e.g. disease history, lab results, procedures), unstructured information remains ubiquitous especially in the context of progress notes and radiology reports. While existing machine learning approaches, such as Natural Language Processing (NLP), make it possible to extract and retrieve relevant information from medical records, they are far from complete, and unsuited for rapid large scale medical record querying and retrieval \cite{Xiao2018}. 

Additionally, medical data often mirrors the underlying disease distribution of a population, reflecting the marked imbalances in the incidence and prevalence rates of many illnesses. This under-representation of certain diseases in medical education as a result of low prevalence has many downstream consequences, resulting in substantial contributions to `miss' errors in screenings and diagnosis \cite{Evans2013}. Moreover, the use of medical data for research and education comes with its own set of privacy and legal hurdles: the growing availability of electronic medical records affords researchers and educators a range of opportunities, at the cost of growing ethical issues, ranging from debates surrounding the quality of de-identification \cite{Moore2015} to ongoing discussions on data ownership, access and control.

In this paper, we propose the use of generative adversarial networks (GANs) as a primer for education in the field of radiology. We train a style-based GAN architecture developed by Karras et al. \cite{Karras2020AnalyzingAI} on an in-house dataset of mammograms collected from the American University of Beirut Medical Center (AUBMC), with approval from the Institutional Review Board (IRB). We then demonstrate the controlled modification of global and local image attributes in the generated images to obtain mammograms with specific characteristics of interest. A double-blind study on four breast radiologists is then performed to assess the visual quality of the resulting images. Finally, we discuss the limitations of our methodology, and provide possible applications for use in clinical settings.
% , and suggest improvements to enhance on the quality of the generated images.

\section{Background}
\subsection{Generative Adversarial Networks}
Generative Adversarial Networks (GANs), proposed by Goodfellow et al. \cite{Goodfellow2014GenerativeAN}, are part of a subset of machine learning algorithms known as generative models that enable the generation of new data points by closely matching the underlying distribution of a dataset. In essence, a GAN pits two neural networks, a generator \textit{G} and a discriminator \textit{D}, against each other: the generator must synthesize data in such a way that the discriminator cannot distinguish the real data points from the synthetic ones produced by the generator. In other words, \textit{D} and \textit{G} engage in a min-max game with the following value function $V(D,G)$:

\begin{align*} 
\min_G \max_D V(D,G) = \; &\mathbb{E}_{x \sim p_x}[\log{D(x)}]\;+ \\ &\mathbb{E}_{z \sim p_z}[\log{(1-D(G(z)))}],
\end{align*}

where $x$ is a `real' sample from the actual dataset, represented by distribution $p_x$, and $z$ is a `latent vector' sampled from distribution $p_z$, typically noise.

\begin{figure*}
\centering
  \includegraphics[width=\textwidth ]{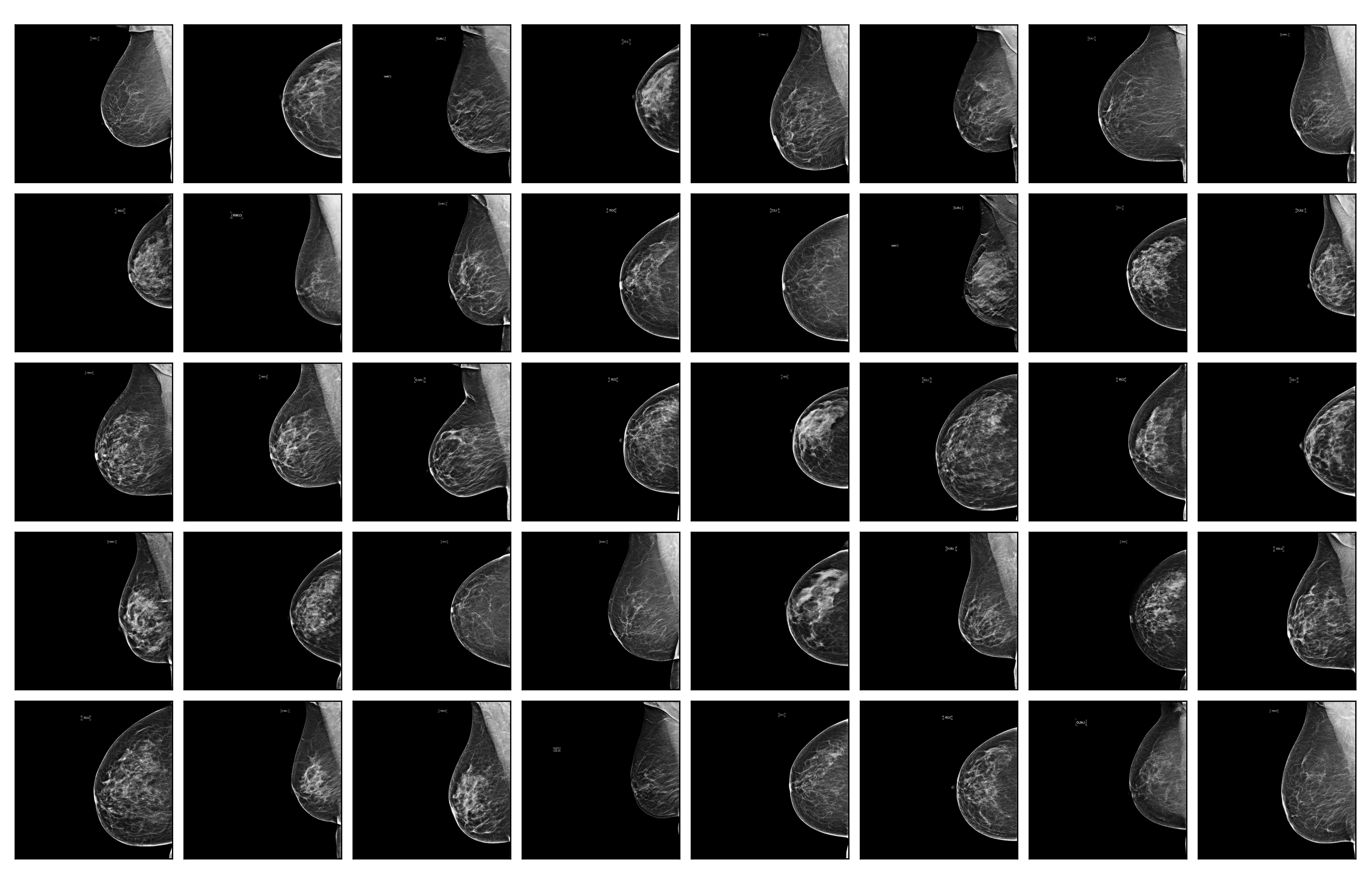}
  \caption{Randomly sampled mammograms generated using a pre-specified random seed and a truncation value of 0.65. }
\label{Figure: main}
\end{figure*}

Since their advent in 2014, the use of GANs for image synthesis has seen a steady increase in the literature, with the introduction of many key innovations to improve on the quality of the generated images and their manipulations. For a thorough review of the GAN literature we refer the reader to recent surveys in \cite{DBLP:journals/corr/abs-1710-07035}.

StyleGAN \cite{DBLP:journals/corr/abs-1812-04948} and the more recent StyleGAN2 \cite{Karras2020AnalyzingAI} are GANs that attempt to learn \textit{disentangled representations} of images, or rather, representations of images with the ability to consistently manipulate the appearance of a semantic attribute in a generated image, independently of any other attribute.

StyleGAN's generator has two sub-networks, one for mapping and one for synthesis. First, the mapping sub-network $M$ maps the input $z$ into an intermediate latent vector $w$. This can be modeled by the following function: 
\begin{align*} 
y_i = G_i(y_{i-1}, w) \textrm{ with } w = M(z)
\end{align*}
with $M$ being an 8-layer multilayer perceptron, and $y_i$ being the input to subsequent layers in the network. The resulting vector is injected as input to all intermediate layers of the synthesis sub-network (G). The authors demonstrate that by allowing each layer to have its own $w_i$, better disentanglement properties are achieved \cite{DBLP:journals/corr/abs-1907-10786}.

\begin{figure*}
\centering
  \includegraphics[width=\textwidth ]{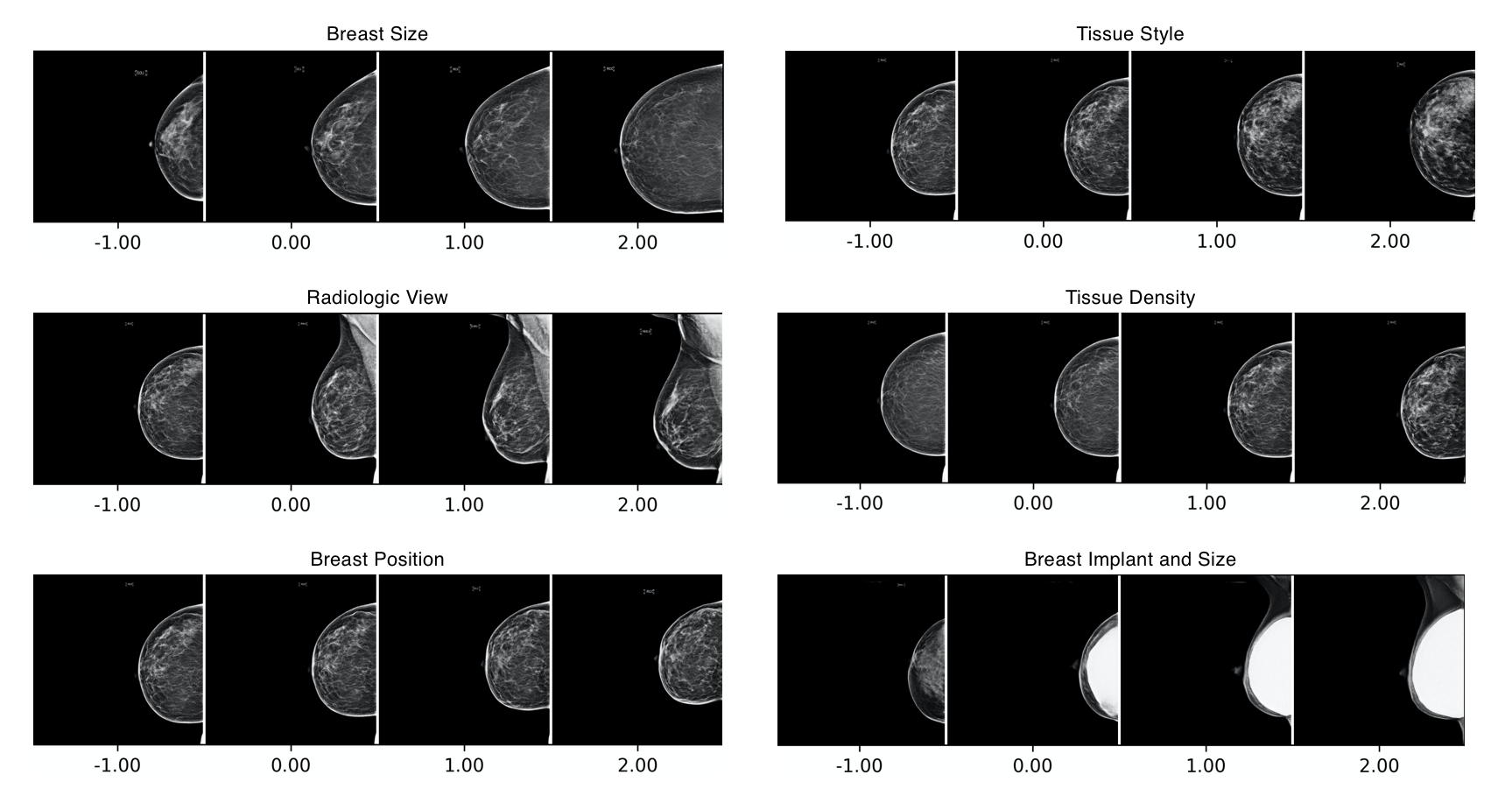}
  \caption{Uncurated rows of global edits which illustrate the six largest principal components in the intermediate W latent space of StyleGAN2, which span the major variations expected of mammograms such as radiologic view and breast density. At each row, the image at 0.00 is the original image along the edit direction.}
 \label{Figure: 1}
\end{figure*}

It is important to note that despite all of the recent improvements made to GANs, \textit{mode collapse} is a phenomenon that has often been described in the context of GAN training. Simply, the generator fails to learn the underlying distribution of the data and only produces a limited variety of samples \cite{arora2018do}. While some improvements to the loss function of the GAN (e.g. Wasserstein loss, spectral regularization) have been shown to reduce the likelihood of this phenomenon, caution should be exercised when using GANs in clinical settings, especially for tasks such as data augmentation in machine learning where mode collapse could lead to the amplification of underlying biases or unwanted distribution shifts.

\subsection{Semantic Editing of Generated Images}
Recent advancements in generative networks have focused primarily on making improvements to the quality of the generated images, and have provided little in terms of controlling the generated outputs. Network architectures like StyleGAN and Conditional GANs, while offering the ability to transfer style vectors or sample from different image classes \cite{DBLP:journals/corr/MirzaO14}, are still limited in terms of the extent of their editing abilities.

For this reason, several works have explored different methods for semantic image editing, ranging from activation-based techniques to latent code-based approaches.

Activation-based techniques directly manipulate activation tensors at specific layers of the generator to modify an image. While previous works have shown success in this direction, they often require some form of supervised learning, which can be difficult and expensive, especially when it comes to medical imaging. For this reason, Härkönen et al. \cite{Hrknen2020GANSpaceDI} demonstrate the use of Principal Components Analysis (PCA) in the activation space of specific layers, allowing high-level control over image attributes without any supervision. In short, PCA analysis is performed on $w_i = M(z_i)$ values of $N$ randomly sampled $z_{1:N}$ vectors to obtain a matrix $V$. Given a new image defined by $w$, emerging high-level attributes be edited by varying the PCA coordinates of $x$:
\begin{align*} 
w^{\prime} = w + Vx,
\end{align*}
where the individual entries $x_k$ of $x$ correspond to different edits, many of which control mostly large-scale variations in the images.

\begin{figure*}
\centering
  \includegraphics[width=\textwidth ]{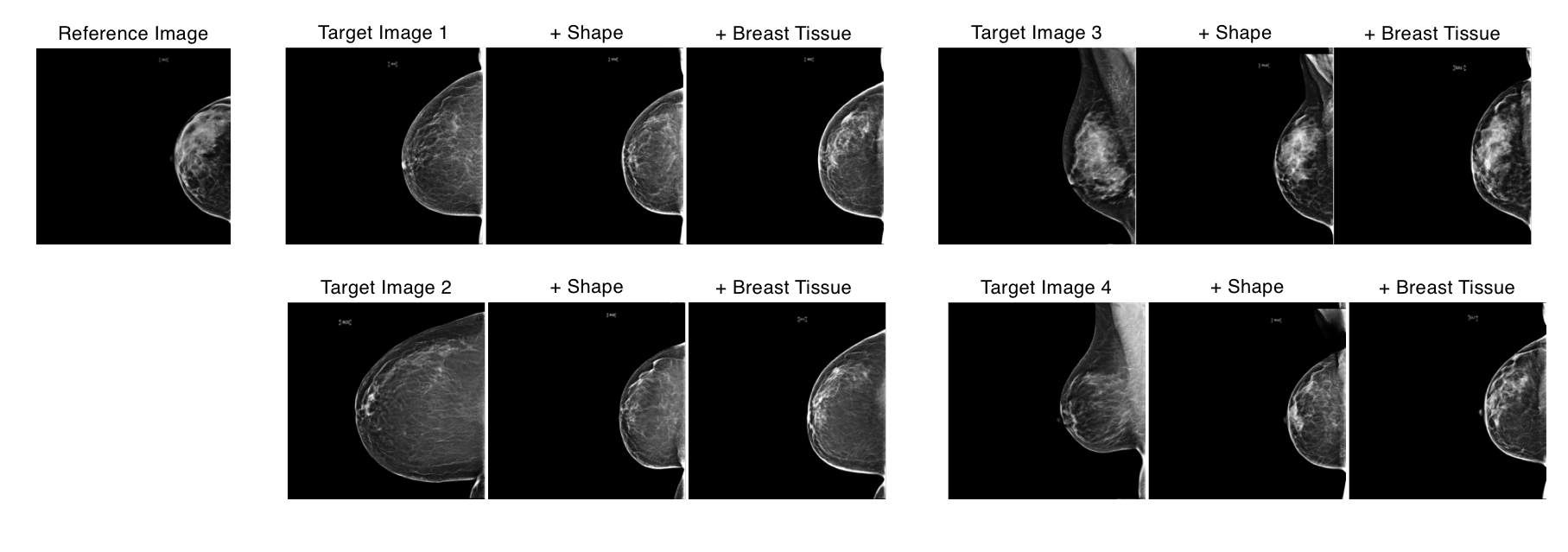}
  \caption{Characteristics of interest such as breast shape and tissue are transferred to the target images by primarily affecting the region of interest. This method allows necessary global changes to the resulting image in order to preserve anatomy and photorealism.}
 \label{Figure: 2}
\end{figure*}

On the other hand, latent code-based techniques learn a manifold in latent space, and perform semantic edits by traversing paths along this manifold \cite{DBLP:journals/corr/PerarnauWRA16};\cite{DBLP:journals/corr/ZhuKSE16}, usually making use of heavy external supervision. In their seminal paper, Collins et al. \cite{Collins2020EditingIS} propose a latent code-based approach that permits local edits by borrowing attributes of interest from reference images. This is achieved by applying spherical k-means clustering along the activation layers of a given layer of a trained generator to obtain clusters corresponding to semantic parts of an image \cite{Collins2018DeepFF}. A simplified understanding of the conditional interpolation of the feature of interest can then be modeled by:
\begin{align*} 
i^{\prime} = s + Q(r - s) 
\end{align*}
with $I^{\prime}$ being the generated image, $S$ and $R$ being the target and reference images respectively, and $Q$ being the query vector, a diagonal matrix containing the values of a semantic cluster controlling a region of interest, along with a function to control the strength of the interpolation.

\subsection{Motivation and Related Work}
Modern Electronic Health Records (EHRs) provide troves of data ripe for use in human and machine learning. However, this opportunity presents itself with a new set of challenges and limitations.

Despite the growing number of medical imaging performed each year, it is frequently the case that medical datasets suffer from severe class imbalances, along with incompletely annotated or insufficient data \cite{DBLP:journals/corr/abs-1806-00388} \cite{pmid29618526}. Images are often accompanied by unstructured data with language irregularities and ambiguities, that complicate the use of common machine learning methodologies. This makes it difficult to query and fetch relevant data pertaining to a specific disease or its clinical presentation. 

Additionally, medical datasets commonly exhibit strong class imbalances and bias. While healthy individuals might be underrepresented in hospital settings, the opposite is also true for most screening programs, including mammography. For example, the prevalence of breast cancer in a screening population is often cited as laying between 0.5 and 1.0\% (Global Burden of Disease Study, 2017). With the inclusion of both standard views (CC and MLO) for each breast in a dataset, along with the observation that malignancies in both sides is relatively rare, it is possible that as many as 99.7\% of the images will be benign \cite{ranschaert:2019}. These imbalances coupled with the privacy and legal constraints surrounding work with sensitive health records, and a relative inability to freely share them across institutions,  make it difficult in many cases for individual researchers and medical trainees to compile sufficient examples for human or machine learning tasks.

A number of attempts have been made to apply GANs to medical imaging datasets \cite{DBLP:journals/corr/NieTPRS16}\cite{DBLP:journals/corr/abs-1708-01155}\cite{DBLP:journals/corr/abs-1807-10225}\cite{Sandfort2019} ranging from the generation of synthetic MRI images with brain tumors, to the modification of contrast CT images for segmentation. Moreover, several approaches have demonstrated the feasibility of generating high-resolution mammograms for use in data augmentation and domain transfer \cite{korkinof:MIDLAbstract2019a} \cite{8621223}. In 2018, Finlayson et al. \cite{DBLP:journals/corr/abs-1812-01547} proposed the idea of utilizing GAN generated images for radiology education but only demonstrated that the trained GAN had learned clinically-relevant features sufficient for machine learning. 

In this paper, we demonstrate the possibility of utilizing GANs as a source of training for humans through mammogram generation, and reveal global and local editing capabilities for the modification of attributes of interest, to provide an exceptionally large set of examples for human training and visualization of common and rare breast pathologies.

\section{Methods}
IRB approval was granted for all stages of this study. A dataset of 162,988 mammograms consisting of the four standard views used in breast cancer screening (R-CC, L-CC, R-MLO, L-MLO) was collected from (AUBMC) for all women aged 18+ between the dates of January 1, 2012 and September 05, 2019.

Data preprocessing consisted of several steps. In order to speed up computations and work within the scope of the available hardware resources, all mammograms were first downsized to a height of 512 pixels before appending black pixels to the edge of the mammogram opposing the breast to obtain square images of shape (512, 512). Mammograms were then flipped along the vertical axis in order to align all of the breasts and increase StyleGAN training stability. The dataset was then split into 152,973 images for training and 10,015 images for testing.

Several initial experiments were first conducted to assess the scope of the study, and to determine best practices. The model was initially trained on low-resolution images, with all breasts centered along their horizontal axes of symmetry, the latter of which yielded sub-optimal results due to the inevitable cropping of lower and apical breast tissue.

The final model's implementation, leveraged from Karras et al. 2019 with hardware-specific modifications and some memory optimizations to account for the size of the dataset, was trained for 4 days on a single Tensor Processing Unit (TPU) v3 to synthesize high-resolution (512 x 512) mammograms, for a total of 10,000,000 images exposed to the discriminator.

% \begin{figure*}
% \centering
%   \includegraphics[width=\textwidth ]{images/metrics.png}
%   \caption{Characteristics of interest such as breast shape and tissue are transferred to the target images by primarily affecting the region of interest. This method makes necessary global changes to the resulting image in order to preserve anatomy and photorealism.}
% \end{figure*}

\begin{table*}[ht]
\setcounter{table}{1}

\begin{subtable}{\linewidth}
\centering
\begin{tabular}{P{0.20\linewidth}P{0.20\linewidth}P{0.20\linewidth}}
\hline
Radiologist & AUC & Precision\\
\hline
1 & 0.56 & 0.58\\
2 & 0.74 & 0.78\\
3 & 0.45 & 0.42\\
4 & 0.39 & 0.41\\
  &      & \\
Average & 0.54 & 0.55\\
\hline
\end{tabular}
\caption{Binary Classification Task}
\label{tab:classification}
\end{subtable}
\hfill{}

\vspace*{0.5 cm}

\begin{subtable}{\linewidth}
\centering
\begin{tabular}{P{0.20\linewidth}P{0.20\linewidth}P{0.20\linewidth}}
\hline
Radiologist & Time per round (in seconds) & Total Number of Rounds \\
\hline
1 & 40.1 & 3\\
2 & 24.6 & 14\\
3 & 24.0 & 3\\
4 & 14.0 & 3\\
  &   &  \\
Average & 25.7 & 5.75\\
\hline
\end{tabular}
\caption{Discrimination Task}
\label{tab:out}
\end{subtable}

\caption{Performance of four expert radiologists on different classification tasks on a dataset composed of 100 real, synthesized and edited mammograms.}
\label{Table 1}
\end{table*}

\section{Experiments}
\subsection{Global Editing of Attributes}
To carry out global edits of generated mammograms, PCA analysis was conducted in the intermediate $W$ latent space of StyleGAN2 using 100 components. Random components were visually inspected by modifying edit directions and constraining the variations to only a subset of layers, while leaving other layers’ inputs unchanged. For truncation, we use a value of 0.65, since greater values tended to produce anatomically distorted mammograms.

\subsection{Local Editing of Attributes}
Local editing of attributes of interest was achieved by performing spherical $k$-means clustering with k=10 on the first 8$\times$8 resolution layer of the generator. We set $\rho$ such that $\frac{\rho}{1+\rho} = 0.1 $ and tune $20 \leq \epsilon \leq 100$ for best performance based on the target image and region of interest. The process of qualitative evaluation required only minutes of human supervision.

\subsection{Visual Turing Test}
A double-blind study on four expert mammography radiologists was carried out in order to determine the quality of the generated images. The radiologists have all previously completed a fellowship in breast radiology, and average more than 8 years of clinical experience.

The dataset was created by first sampling more than 1000 images from our GAN with a truncation value set at 0.65, and applying a random edit to the generated images with a probability $\epsilon=0.35$. An image is then sampled without replacement at random from either the test set of real radiographs or the pool of synthesized and edited mammograms to obtain a dataset of 100 images. Edits consisted of semantic changes to breast size and shape, tissue density, radiologic view, breast position, as well as the presence and size of implants (Figs.\:\ref{Figure: 1}, \& \ref{Figure: 2}). The final composition of the dataset is summarized below:

\begin{table}[ht]
\setcounter{table}{0}
\begin{subtable}{\linewidth}
\centering
\begin{tabular}{P{0.28\columnwidth}P{0.28\columnwidth}P{0.28\columnwidth}}
\hline
Real & Synthesized & Edited\\
\hline
52 & 31 & 17\\
\hline
\end{tabular}
\caption{Final Dataset Composition}
\label{Table 2a}
\end{subtable}
\hfill{}

\vspace*{0.5 cm}
\begin{subtable}{\linewidth}
\centering
\begin{tabular}{P{0.45\columnwidth}P{0.45\columnwidth}}
\hline
Radiologist & Post-Fellowship Years of Experience\\
\hline
1 & 18\\
2 & 3\\
3 & 11\\
4 & 2 \\
\hline
\end{tabular}
\caption{Radiologist Statistics}
\label{Table 2b}
\end{subtable}
\caption{Radiologist and dataset summary statistics}
\end{table}

In the first task framed as a binary classification problem, the radiologists were presented with a series of images from the dataset and were tasked with classifying each image as real or generated. Performance was measured after the classification of all of the images in the dataset.

In the second task, the radiologists were presented with six random images simultaneously (five real and one synthesized mammogram), and were tasked with determining the generated image in each round. Performance was measured after three wrong answers.

\section{Results}
Figure \ref{Figure: main} depicts forty randomly sampled mammograms from the GAN, along with disentangled global and local edit directions in figures \ref{Figure: 1} and \ref{Figure: 2} respectively. Visual inspection of the mammograms reveals a variety of styles and pathological features with no signs of obvious mode collapse.

Results of the Visual Turing Test experiments are presented in Table \ref{Table 1}. For the binary classification task (Table \ref{tab:classification}), the radiologists are no better than random at differentiating between synthesized and real mammograms, with an average AUC of 0.54 and a precision of 0.55. While radiologist 2 manages to obtain an AUC of 0.74, performances of radiologists 3 and 4 hover below that of a random classifier defined as a performance with an AUC of 0.5. Radiologist 1 performs only slightly better than random with an AUC of 0.56.

In the discrimination task (Table \ref{tab:out}), the radiologists struggled to identify the generated mammograms, with average performance lasting around 5.75 rounds. While radiologist 2 managed to reach fourteen rounds, the remaining radiologists never made it past the third round. 

\section{Discussion}
Based on the Visual Turing Test results obtained, it is clear that the generative network has learned important visual features that enable it to generate mammograms indistinguishable from real ones at this resolution, even to expert radiologists. On average, classification performance is no better than random, with successful differentiation achievable mostly after lengthy deliberation. However, it is important to keep in mind that while experiments were carried out at resolutions of 512x512, radiologists typically work with mammograms exceeding 3000 pixels in both dimensions, and that classification and discrimination performances were obtained for only a small sample of radiologists. While previous works \cite{korkinof:MIDLAbstract2019a} have successfully reported generating high-quality mammograms at greater resolutions, further work is needed to evaluate the quality of local and global editing operations to resolutions greater than 512 pixels.

Additionally, these methods of generation and editing pose some interesting challenges. While some attributes, such as the presence or absence of calcification, are strictly discrete, a learned latent space is usually continuous by nature, resulting in some generated images with attributes that lie in-between the discrete values. The resulting image may not truly reflect accurate pathophysiology or may even reveal some visual inconsistencies, such as the textual radiologic view labels on some mammograms that appear to be halfway between `CC' and `MLO'.

Moreover, despite demonstrating intrinsic disentangled semantic properties, global and local editing operations sometimes require supervised curation of the generated images, as the quality of the results depends heavily on the extent to which an object’s representation is disentangled from other representations. Per-image fine-tuning of edit parameters is also sometimes necessary to obtain optimal results after local editing.

Despite these minor drawbacks, it is clear that image generation will play an important role in future clinical education. Medical GANs provide an opportunity to improve visual diagnosis through the generation of virtually unlimited, high-quality training examples, especially in the case of rare pathologies. The extension of these semantic editing operations to real patient mammograms through real image mapping in latent-space (Abdal et al. 2019) could allow for the modification of clinical attributes in real-time, for use in interactive learning experiences and prototyping such as cancer growth simulation, visualization and privacy preservation.

\section{Acknowledgements}
This research was made possible thanks to important contributions from Shawn Presser and Aydao, as well as the computational resources provided by Google's TensorFlow Research Cloud (TFRC).

\addtolength{\textheight}{-12cm}   % This command serves to balance the column lengths
                                  % on the last page of the document manually. It shortens
                                  % the textheight of the last page by a suitable amount.
                                  % This command does not take effect until the next page
                                  % so it should come on the page before the last. Make
                                  % sure that you do not shorten the textheight too much.

%%%%%%%%%%%%%%%%%%%%%%%%%%%%%%%%%%%%%%%%%%%%%%%%%%%%%%%%%%%%%%%%%%%%%%%%%%%%%%%%

\medskip

\printbibliography
\end{document}